\documentclass[10pt,tightenlines,showpacs,letterpaper,aps,prd,longbibliography,onecolumn,nofootinbib,superscriptaddress]{revtex4-2}
\usepackage{amssymb}
\usepackage{epstopdf}
\usepackage{amsmath,amsthm}	
\usepackage{bm}
\usepackage{etex}
\usepackage{hyperref}
\usepackage{amsfonts}
\usepackage{physics}
\usepackage{graphicx}
\usepackage{xcolor}
\usepackage{multirow}
\begin{document}
	
	\title{On the Fidelity Robustness of CHSH--Bell Inequality via Filtered Random States}
	
	\author{Antonio Mandarino}
	\affiliation{International Centre for Theory of Quantum Technologies (ICTQT), University
		of Gdansk, Wita Stwosza 63, 80-308 Gda\'{n}sk, Poland}
	\author{Giovanni Scala}
	\affiliation{International Centre for Theory of Quantum Technologies (ICTQT), University
		of Gdansk, Wita Stwosza 63, 80-308 Gda\'{n}sk, Poland}

	\begin{abstract}
		\noindent The theorem developed by John Bell
		constituted the starting point of a revolution that translated a philosophical question about the nature of reality into the broad and intense field of research of the quantum information technologies. 
		We focus on a system of two qubits prepared in a random, mixed state, and we study the typical
		behavior of their nonlocality via the CHSH--Bell inequality.
		Afterward, motivated by the necessity of accounting for inefficiency in the state preparation,
		we address to what extent states close enough to one with a high degree of nonclassicality can violate local realism with a previously chosen experimental setup.
	\end{abstract}
	
	\maketitle
	\section{Introduction}
	John Bell introduced in 1964 the inequality that is known today as a tool to test the predictions of quantum mechanics in contrast to those that one can obtain in any classical scenario involving local hidden variable (LHV) models \cite{Bell_original}. However, the most experimental-friendly generalisation of their inequality was formulated years later by Clauser--Horne--Shimony--Holt (CHSH), and it was immediately applied to check whether the predictions of quantum mechanics are in agreement with the bounds set by the inequality. As a matter of courtesy in recognising the work of all the authors but to emphasise the novelty in the idea of Bell, we will refer to such a theorem in the form of an inequality to the CHSH--Bell theorem.
	
	The first experimental tests in favour of the quantum mechanical predictions were conducted by Freedman and Clauser in 1972 \cite{Freedma72},
	in which the authors already noticed that the \emph{``data, in agreement with quantum mechanics, violate these restrictions to high statistical accuracy, thus providing strong evidence against local hidden-variable theories''}.
	Their experimental platform was based on the measurement of photon linear momentum and was an optical realisation of the scheme proposed by Bohm \cite{Bohm_1, Bohm_2} to reformulate the famous EPR paradox~\cite{EPR} in terms of a pair of spin-1/2 particles.
	However, several Bell tests have been performed to verify the Bell theorem \cite{Freedma72,Aspect1982,Weihs1998,Rowe2001,Matsukevich2008,Ansmann2009,Scheidl2010,Hofmann2012} and the final ones closing all the loopholes affecting the detection and locality assumptions were only recently performed \cite{loophole_1, loophole_2, loophole_3}.

	In this paper, we address to what extent a typical two-qubit state described by a random density matrix violates (or not) the local realism via means of the deviation from the CHSH--Bell bound.
	Specifically, we studied the geometrical properties of the Hilbert space by showing that even if a given qubit--qubit state $\rho$ drastically violates the classical bound, such violation vanishes for some states in a very narrow neighbourhood of $\rho$ that we quantify by the fidelity. In this manuscript, we quantify how narrow the neighbourhood must be such that the states within violate the classical bound. This is a crucial aspect because nowadays the Bell--CHSH theorem is also used to certify quantumness by self-testing~\mbox{\cite{Coopmans2019,Supic2020,Gigena2022}}.
	
	{In particular, we focused on the class of state generated in a way that their marginal distributions are the maximally mixed states as we will clarify later on. This procedure increases the detectability of nonclassical features, as, for instance, in entanglement detection.
		It is well known 
		that local operations do not vary the amount of entanglement in a quantum state, but they can increase the detection efficiency via an entanglement witness~\mbox{\cite{Guehne2007,Guehne2010}.}}
	
	Despite their simplicity, the tools developed to assess the violation of local realism for the simplest example of multipartite quantum systems, namely a two-qubit one,
	have been recently proven to constitute a powerful apparatus to {assess} more complicated states such as bosonic fields, for which the number of particles is undefined
	\cite{1stPaper,1stPLA,2ndPaper,CommentDun,Non_contexAM}.
	
	This paper is organised as follows: {In Section \ref{sec:appendix_random}, we explain how to randomly generate the filtered quantum states.} In Section \ref{sec:Bell}, we review the formalism needed to derive the CHSH--Bell inequality and address the typicality for random mixed two-qubit states. Section \ref{sec:fidelity} is devoted to a discussion of fidelity, a measure used for quantum technology purposes to determine the ``closeness'' of two quantum states in the Hilbert space, and we describe how it relates to the assessment of Bell nonclassicality when an imperfect state preparation is taken into account.
	
	\section[\appendixname~\thesection]{Random Density Matrices}
	\label{sec:appendix_random}
	An easy way to generate a random bipartite quantum state can be pursued by borrowing the notion of
	the Schmidt decomposition of a vector for unitary matrices.
	A unitary matrix $U$ of dimension $N^2$ is ``vectorised'', such that it constitutes
	an element of the vectorial Hilbert--Schmidt space of matrices \cite{unitary_karol,Unitary_weyl}.
	Hence, it can be Schmidt decomposed as~follows:
	\begin{equation}
		\label{genU}
		U= N \sum_{i=1}^{N^2} \sqrt{\lambda_i} \; A_i \otimes B_i.
	\end{equation}
	{The matrices} 
	$A_i$  and $B_i$ (in general not unitary) are two orthonormal bases in the space
	of operators and satisfy the following relations
	$\mathrm{Tr}{A_i^\dag A_j}= \mathrm{Tr}{B_i^\dag B_j} = \delta_{ij} $.
	As usual, we can associate to such matrix $U$ a Schmidt vector whose components are the expansion coefficients, and
	that is invariant with respect to local unitary operations.
	
	Moreover, any matrix $O$ acting on a bipartite  $N \times N$ system can be expanded over the
	product basis $\ket{m} \otimes \ket{\mu},$ so for shorthand notation, we label the matrix element as
	$O_{\stackrel{\scriptstyle m \mu}{n\nu}} = \bra{m \mu}O \ket{n  \nu}.$
	For any such $O$ we define the realigned matrix $O^R$,
	obtained \emph{from blocks to rows} by reshaping its square blocks of order $N$ into rows of length $N^2$,
	with entries $ O_{\stackrel{\scriptstyle m \mu}{n\nu}}^R =O_{\stackrel{\scriptstyle m n}{\mu \nu}}${, i.e.,
		\begin{equation}
			\left(\begin{array}{cc|cc}
				o_{11} & o_{12} & o_{13} & o_{14}\\
				o_{21} & o_{22} & o_{23} & o_{24}\\
				\hline o_{31} & o_{32} & o_{33} & o_{34}\\
				o_{41} & o_{42} & o_{43} & o_{44}
			\end{array}\right)\xrightarrow{R}\left(\begin{array}{cccc}
				o_{11} & o_{12} & o_{21} & o_{22}\\
				\hline o_{13} & o_{14} & o_{23} & o_{24}\\
				\hline o_{31} & o_{32} & o_{41} & o_{42}\\
				\hline o_{33} & o_{34} & o_{43} & o_{44}
			\end{array}\right).
		\end{equation}
	}Then, the Schmidt vector of the unitary matrix U is equal to the eigenvalues of a positive~matrix
	\begin{equation}
		\label{eq:resh}
		\rho = \frac{1}{N^2}U^R (U^R)^\dag
	\end{equation}
	where $U^R$ stands for the realigned matrix \cite{karol_geometry}.
	
	We note that the positive random  matrix obtained by multiplying the realignment unitary matrix with its Hermitian conjugate,
	$U^R (U^R)^{\dagger},$ behaves as a Wishart matrix and hence the probability distribution of eigenvalues
	of $\rho$ reduces to the Hilbert--Schmidt measure \cite{zyczkowski2011generating, enriquez2015minimal, enriquez2018entanglement}.
	
	In particular, for our purposes, we generated a set of random unitary matrices from the circular unitary ensemble (CUE) that consists of all unitary matrices distributed according to the normalised Haar measure on the unitary group $U(N)$
	\cite{zyczkowski1994random}.

	\section{The CHSH--Bell Inequality }
	\label{sec:Bell}
	A bipartite Bell--CHSH scenario consists of two physical systems, traditionally called Alice and Bob, spatially separated, in the sense that, there exists a reference frame where Alice and Bob's measurements are simultaneous events located in different places \cite{gigena2022a}. In particular, both Alice and Bob simultaneously choose, respectively, $x\in\{1,2\}$ and $y\in\{1,2\}$ to perform, at a time, one out of two different measurements, $A_1$, $A_2$, and $B_1$, $B_2$. Each binary measurement yields two possible outcomes, let us say $\{-1,+1\}$ associated with POVM elements $A_x=\{A_x^{-1},A_x^{+1}\}$, and $B_y=\{B_y^{-1}, B_y^{+1}\}$, with $x,y\in\{1,2\}$, respectively, for Alice and Bob.
	
	Therefore, according to Born's rule, it is possible to compute the probabilities for each possible event. In fact, we will say that
	{More specifically, a rigorous mathematical notation would describe the event as an element of a sigma algebra $\{A_x=a, B_y=b\}$ and the probability that it happens is $p\{A_x=a, B_y=b\}\equiv p(a,b|x, y)$, where $p(\cdot|\cdot)$ is not a conditional probability. We keep the physics notation according to the tradition in the related literature.}
	The probability that the outcome of Alice is $a\in\{-1,+1\}$ and the outcome of Bob is $b\in\{-1,+1\}$ given that Alice chooses to measure $A_x$ and Bob $B_y$ is
	\begin{equation}\label{Born}
		p(a,b|x, y)=\mathrm{Tr}\left(
		\rho A_a^x\otimes B_b^y
		\right).
	\end{equation}
	In particular, to violate the Bell--CHSH scenario, Alice and Bob share an entangled state $\rho\in\mathcal{B}(\mathbb{C}^2\otimes\mathbb{C}^2)$, and the POVM elements $A_a^x$ and $B_b^y$ that act locally on the space of bounded operators $\mathcal{B}(\mathbb{C}^2)$ represent spin--$1/2$ measurements along the directions of $\mathbb{R}^3$ $\hat{\textbf{a}}$, $\hat{\textbf{a}}^\prime$ for Alice and $\hat{\textbf{b}}$, $\hat{\textbf{b}}^\prime$ for Bob.
	Thanks to Naimark's theorem, we can consider without loss of generality that the measurements are projective, i.e., $A_{x}^{k}A_{x}^{k^\prime}=\delta_{k k^\prime}A_{x}^{k}$ and $M_{y}^{k}M_{y}^{k^\prime}=\delta_{k k^\prime}M_{y}^{k}$.
	Therefore, to compute the probability of Equation \eqref{Born} the two dichotomic POVM elements of the binary measure, let us say along the direction $\textbf{n}\in\mathbb{R}^3$, are the projection operators onto the outcome \emph{spin up}, (or $+1$) and spin down, (or $-1$) states, i.e., $M_n^\pm=1/2(\bm1\pm \hat{\textbf{n}}\cdot \bm \sigma)$, for $M=A,B$.
	Since the measurements in this scenario are dichotomic, one can omit the identity~\cite{Braunstein1992} such that the measurements for Alice and Bob are, respectively, $x\in \{\hat{\textbf{a}}\cdot\bm\sigma,\hat{\textbf{a}}^\prime\ \cdot \bm{\sigma}\}$ and $y\in \{\hat{\textbf{b}}\cdot\bm\sigma,\hat{\textbf{b}}^\prime\ \cdot \bm{\sigma}\}$.
	Because of this,  instead of dealing with probabilities as in the Bell framework, we can describe the same physics in terms of the CHSH functional, by using the correlation function
	\begin{equation}
		\langle A_xB_y\rangle=\sum_{a,b\in I} ab \,p(ab|xy),\label{2points}
	\end{equation}
	that take values in $[-1, 1]$. In this way the well-known \emph{Bell--CHSH functional} is constructed:
	\begin{equation}
		\label{eq:Bval}
		\mathcal{B}=
		\hat{\textbf{a}}\cdot \bm\sigma \otimes (\hat{\textbf{b}}+\hat{\textbf{b}}^\prime)\cdot\bm\sigma
		+
		\hat{\textbf{a}}^\prime\cdot \bm\sigma \otimes (\hat{\textbf{b}}-\hat{\textbf{b}}^\prime)\cdot\bm\sigma.
	\end{equation}
	Then {the CHSH--Bell-type inequality} obtained, under the assumption of \emph{Bell locality} and \emph{realism} of pre-existing outcomes, reads
	\begin{equation}
		|\langle \mathcal{B}\rangle_\rho|=|\mathrm{Tr}(\mathcal{B}\rho)|\leq 2.
	\end{equation}
	Now, it is well known that for rank-1 density operator, the problem of violating the Bell-type inequality has been completely solved \cite{Selleri1973,Gisin1991,Gisin1992,Popescu1992}. In the years later, Werner shows that entanglement is only a necessary condition to violate Bell inequality, hence there exist mixed Werner entangled states that admit the LHV model. Besides that, several characteristics have been formulated for the violation of Bell functional using the bipartite mixed state shared by Alice and Bob  \cite{Karczewski2022, gigena2022a}. Here, we put the focus on the neighbourhood of a mixed state that violates {the CHSH Bell-type inequality}. Therefore we show that even if there exists a state with a violation in a neighbourhood of Tsirelson bound $2\sqrt{2}$, its neighbourhood does not preserve a sort of \emph{continuity} violation for \emph{na\"ive} values of fidelity. With the term \emph{continuity}, we mean that if an entangled mixed state $\rho_T$ violates {the CHSH--Bell-type inequality} providing a value $2<\langle \mathcal{B}\rangle_{\rho_T}=B_T<2\sqrt{2}$, a \emph{close} state in a neighbourhood of $\rho_T$, let us say $\rho\in\mathcal{I}_{\rho_T}$ provide a value in a neighbourhood of $B_T$. In the following, we are going to present our method in detail.
	
\begin{figure}[h]
		\includegraphics[width= 0.5 \textwidth]{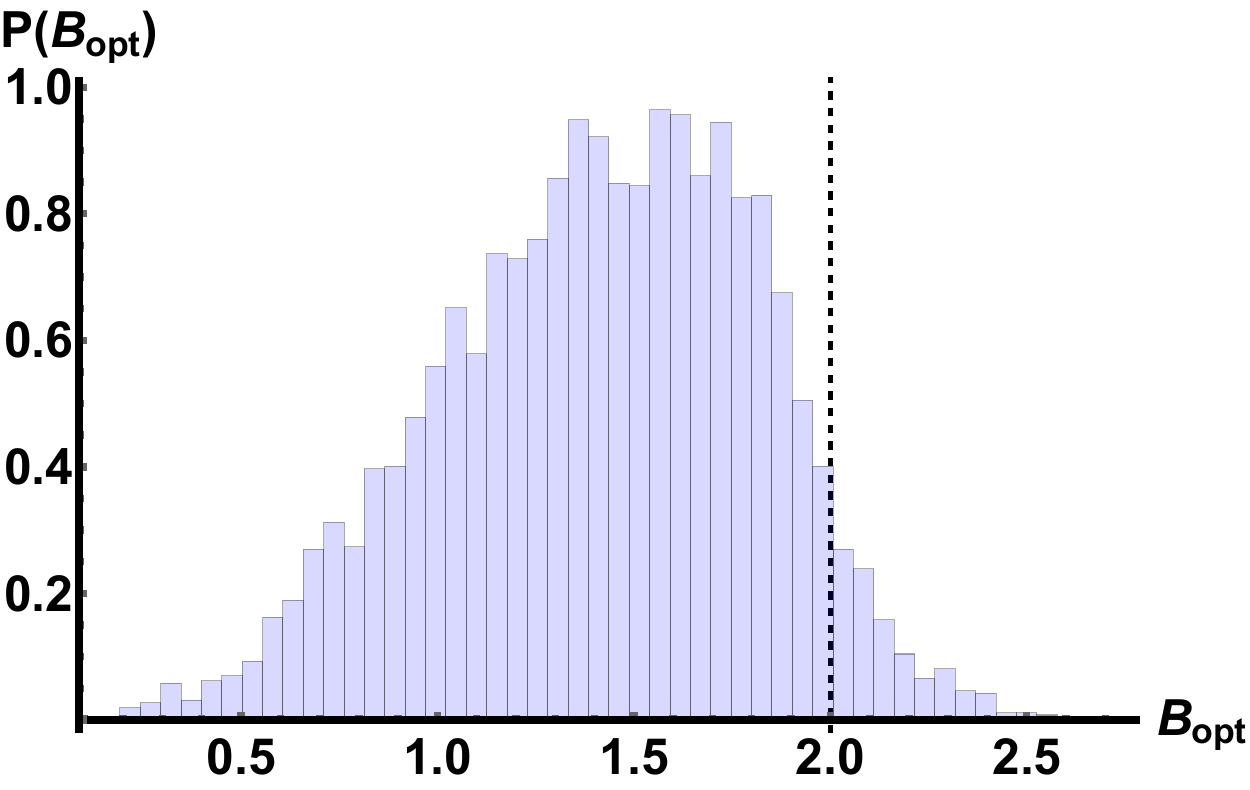}
		\caption{\label{fig:1} The {discrete probability density function $P(B_{\mathrm{opt}})$} of the expression in Equation \eqref{eq:Bopt} for an ensemble of
			{$|\Omega|=5000$} two-qubit random mixed states with the first four central moments $\{1.416, 0.163, -0.251, 2.706\}$.
			The dashed black line at $B_\mathrm{opt}=2$ separates the states admitting a local model $(B_{\mathrm{opt}}\le2)$ and those violating local realism $(B_{\mathrm{opt}}~{>}~2)$.
			The fraction of states leading to a violation of local realism has $p(B_{\mathrm{opt}}>2) \simeq 0.05$2. }
	\end{figure}

	Firstly, we randomly sample a set of mixed states (see {Section \ref{sec:appendix_random} for details}), and for each of them, we optimise the Bell functional with respect to the measurement choices, i.e., the directions $\hat{\textbf{a}},\hat{\textbf{a}}^\prime,\hat{\textbf{b}},\hat{\textbf{b}}^\prime$:
	\begin{equation}\label{eq:dir}
		\max_{\hat{\textbf{a}},\hat{\textbf{a}}^\prime,\hat{\textbf{b}},\hat{\textbf{b}}^\prime}\mathrm{Tr}(\mathcal{B}\rho)\leq 2\sqrt{2}.
	\end{equation}
	The distribution of the state is organised in a normalised histogram in Figure \ref{fig:1}. After that, we select a state, let us call it $ \rho_T$, that gives a value of $|\langle \mathcal{B}_\mathrm{opt}\rangle_{\rho_T}\in [2,72;2.82]$ and sample a neighbourhood \begin{equation}
		\mathcal{I}_{\rho_T}=\{
		\rho \in \mathbb{C}^2\otimes\mathbb{C}^2: \mathcal{F}(\rho,\rho_T) > \alpha
		\},
	\end{equation}
	where $\mathcal{F}(\cdot,\cdot)$ is a notion of \emph{closeness} (see the discussion in Section \ref{sec:fidelity}).
	
\begin{figure}[h]
		\includegraphics[width= 0.32 \textwidth]{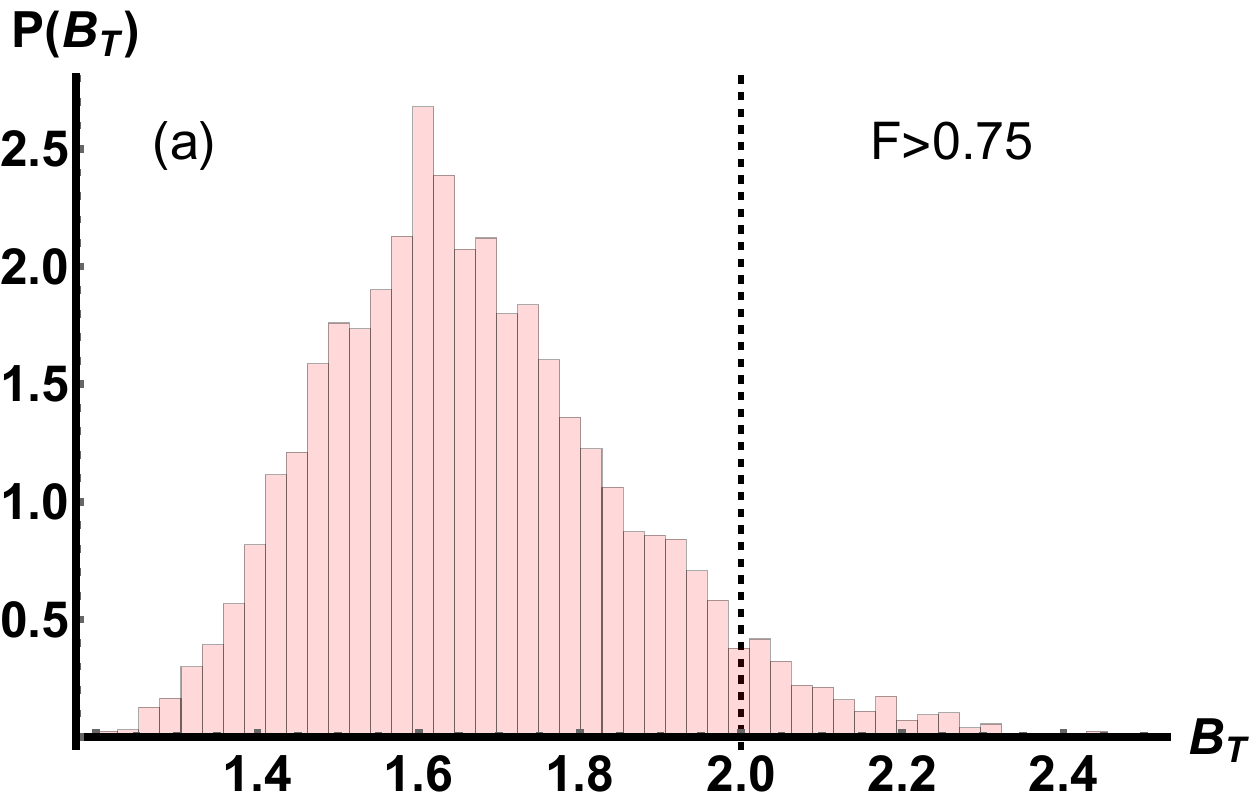}
		\includegraphics[width= 0.32 \textwidth]{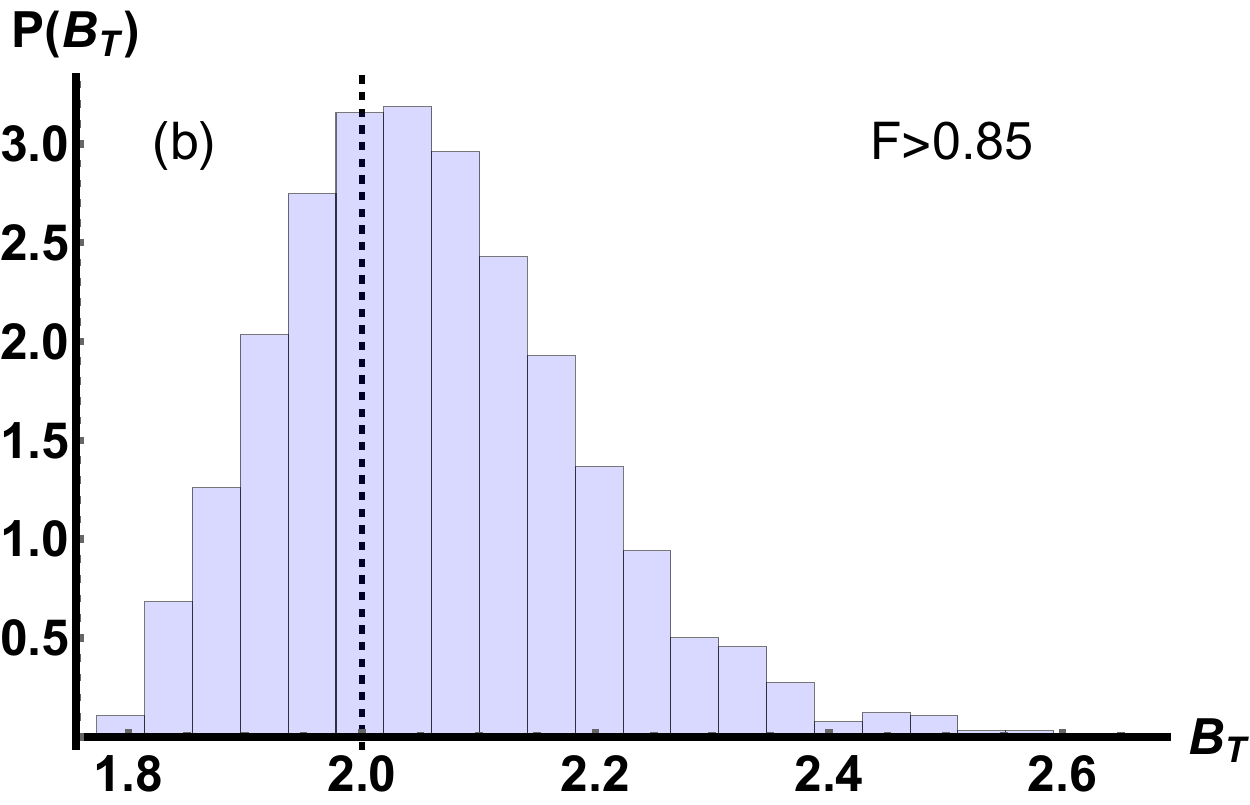}
		\includegraphics[width= 0.32 \textwidth]{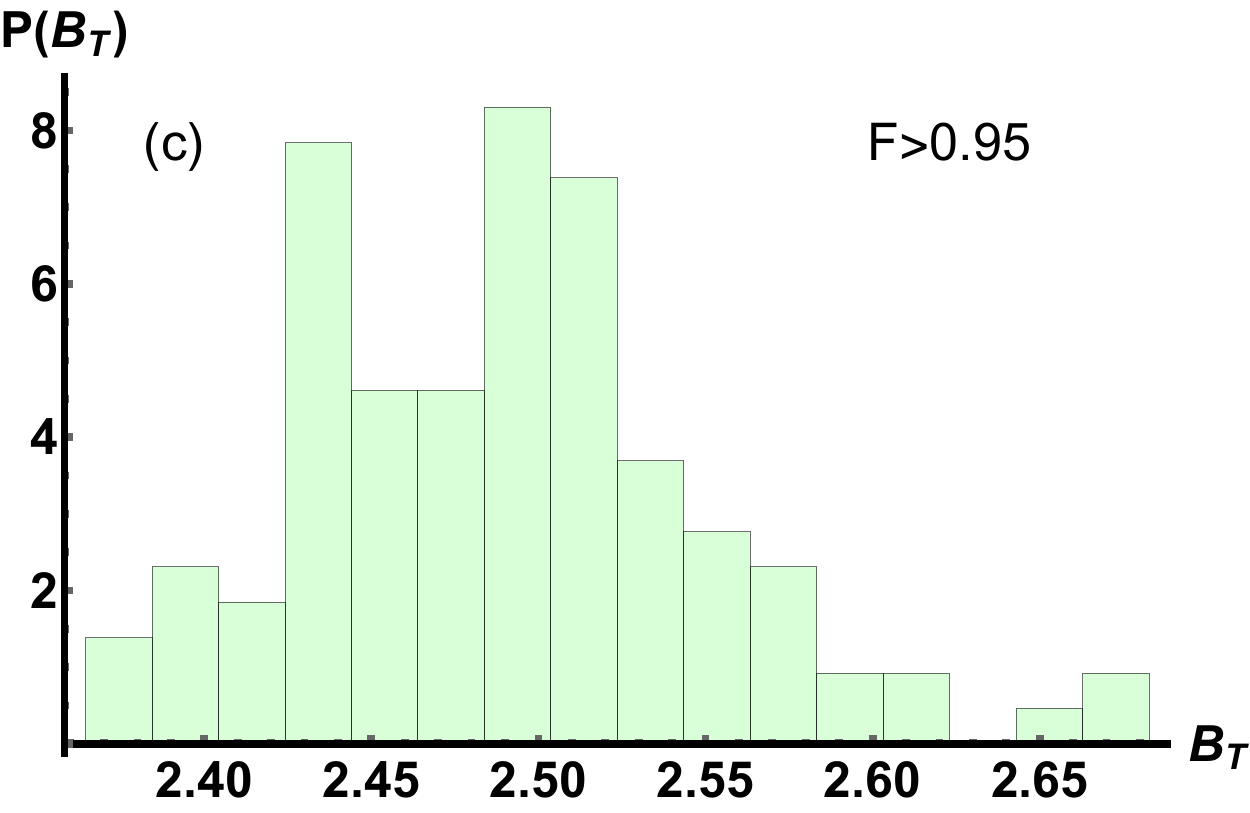}
		\caption{\label{fig:2}
			{The discrete probability density function $P(B_{T})$ of the expression in Equation~\eqref{eq:Bval} for an ensemble of:
				$|F_{\alpha}|=\{4874, 1609, 110\}$ two-qubit random mixed states satisfying the constraint
				$\mathcal{F}(\hat{\rho}_T, \hat{\rho}_s) > \alpha$ with the target state $\hat \rho_T$
				when a fixed choice of measurement settings is considered.
				The corresponding values of fidelity are: in (\textbf{a}) $\alpha=0.75$, in (\textbf{b}) $\alpha=0.85$, (\textbf{c}) $\alpha=0.95$.
				The first four central moments of the three distributions are, respectively:
				(\textbf{a}) $\{1.67, 0.0365, 0.625, 3.518\}$, (\textbf{b}) $\{2.06, 0.0169, 0.737, 3.96\}$, and (\textbf{c}) $\{2.49, 0.0039, 0.699, 3.83\}$.
				The dashed black line demarcates the threshold value separating states admitting a local model $(B_{T}<2)$ and those violating local realism $(B_{T}<2)$.
				The fraction of states leading to a violation of local realism has probability
				(\textbf{a})~$p (B_{T}>2) \simeq 0.058$, (\textbf{b}) $p (B_{T}>2) \simeq 0.65$, (\textbf{c}) $p (B_{T}>2) \simeq 1$. }}
	\end{figure}

	We analyse the histogram shown in Figure \ref{fig:2}, obtained on the states belonging to $\mathcal{I}_{\rho_T}$ to study the \emph{continuity} of  {the CHSH--Bell-type inequality} violation.

	\vspace{-6pt}
	\section{Typicality for the Value of $\mathcal{B}$}
	\label{sec:chsh-random}
	In order to study the typicality of the value assumed by the quantum expectation value of the sum operators defining the CHSH--Bell inequality,
	we generated $M=5000$ random density matrices of dimension $N^2 = 4$ as outlined in Section \ref{sec:appendix_random},
	describing an ensemble of mixed two-qubit states.
	The particularity of such matrices is that they have as marginal states, $\rho_A = \mathrm{Tr}_B{\rho}$ and $\rho_B = \mathrm{Tr}_A{\rho}$, the maximally mixed single qubit state,
	$\rho_A = \rho_B = \bm{1}/2.$
	{The procedure aforementioned is equivalent to generating a density matrix $\rho$ with random entries and applying local operations $(F_A\otimes F_B)\rho(F_A\otimes F_B)^\dagger$ (opportunely normalised) such that $F_A, F_B$ are chosen to obtain $\rho_A = \rho_B = \bm{1}/2.$}
	
	Later on, for any of those states, we have computed the value of the function in Equation \eqref{eq:Bval},
	considering a parametrisation of the general local measurement setting  as
	\begin{equation}
		\hat{\textbf{w}}\cdot \bm\sigma = \left(\begin{array}{cc}
			\cos \theta_w & e^{- i \phi_w} \sin \theta_w \\
			e^{i \phi_w} \sin \theta_w & \cos \theta_w\\
		\end{array}
		\right),
	\end{equation}
	where $\hat{\textbf{w}} = \{ \hat{\textbf{a}}, \hat{\textbf{a'}}\} $ for the first party
	and $\hat{\textbf{w}} = \{ \hat{\textbf{b}}, \hat{\textbf{b'}}\} $ for the second one.
	Eventually, we performed optimisation over eight angles (two angles for each direction)
	{The optimisation algorithm chosen is RandomSearch, which is  a built-in algorithm in the Wolfram
		Mathematica software,  and compared with the other optimisation algorithms offered by the same software, it provides the best results, 
		see also} \cite{SP_3}, 
	and we computed the maximum value; hence, Equation~\eqref{eq:dir} becomes
	\begin{equation}
		\label{eq:Bopt}
		B_{\mathrm{opt}}= \max_{\theta_a, \phi_a,  \theta_a',\phi_a', \theta_b, \phi_b, \theta_b', \phi_b'} |\langle \mathcal{B} \rangle_\rho|.
	\end{equation}
	{Given the sample space of randomly generated quantum states $\Omega\in \mathcal{B}(\mathbb{C}^2\otimes\mathbb{C}^2)$, $B_{\mathrm{opt}}$ is specifically the random variable
		\begin{equation}
			B_{\mathrm{opt}}=\{x\in[0,2\sqrt{2}]\,|\quad \forall\rho\in \Omega: \max_{\theta_a, \phi_a,  \theta_a',\phi_a', \theta_b, \phi_b, \theta_b', \phi_b'} |\langle \mathcal{B} \rangle_\rho|=x\}
		\end{equation}
		with probability distribution $p\{B_{\mathrm{opt}}\}$ displayed in Figure \ref{fig:1}.
	}
	
	It is clear that on average such a set of randomly generate state do not violate the constraints imposed by local realism,
	in fact, we have that $\overline{B_{opt}} = 1.416 \pm 0.404.$

	\section{Fidelity between a High Nonclassical State and a Set of Neighbouring States}
	\label{sec:fidelity}
	Studies devoted to quantifying the similarity between two quantum states have been spurred by advancements in the field of quantum technologies. The Uhlmann Fidelity \cite{uhlmann1976transition} has also gained widespread acceptance among those distance-like values (in addition to measurements that behave as a proper distance in the Hilbert space).
	
	We revise here its main features.
	
	\subsection{Fidelity between Two Quantum States}
	Let be $\hat{\rho}_1$ and $\hat{\rho}_2$, the density matrices of two quantum states the fidelity between them is defined as \cite{uhlmann1976transition}
	\begin{equation}
		\label{Fidelity}
		\mathcal{F}(\hat{\rho}_1, \hat{\rho}_2) =\mathrm{Tr}\left[\sqrt{ \sqrt{\hat{\rho}_1 } \hat{\rho}_2 \sqrt{\hat{\rho}_1
		}}\right]^2\,.
	\end{equation}
	An easy visualisation of the physical meaning of this quantity can be obtained considering pure states.
	In fact if the first state is pure, namely $\hat{\rho}_1 = \ketbra{\psi_1}{\psi_1},$
	the fidelity reduces to $\mathcal{F}({\psi_1}, \hat{\rho}_2)= \bra{\psi_1} \hat{\rho}_2 \ket{\psi_1}$,
	i.e., it reads as the probability to find the state {$\hat{\rho}_2$} in $\ket{\psi_1}.$ The situation becomes even more clear when also the second system is a pure state, hence the Fidelity coincides with the squared overlap between the two states
	$\mathcal{F}(\ket{\psi_1}, \ket{\psi_2})= |\bra{\psi_1} \ket{\psi_2}|^2$.
	It is easy to note that the fidelity is defined in a precise interval,
	i.e., $\mathcal{F}(\hat{\rho}_1, \hat{\rho}_2) \in [0,1]$ with the lower value indicating orthogonal states
	and the upper value for states belonging to the same ray in the Hilbert space.
	
	It is worth remarking that fidelity does not constitute a proper distance in the
	Hilbert space. Despite this observation, it can be put in relationship with proper distances in a Hilbert space
	related to different aspects of the distinguishability of quantum states \cite{Karol_fid}.
	In particular, the Bures distance between two quantum states is defined as a function of the fidelity as
	\begin{equation}
		D_B(\hat{\rho}_1,\hat{\rho}_2)=\sqrt{2[
			1-\sqrt{F(\hat{\rho}_1,\hat{\rho}_2)}]}\,.
	\end{equation}
	Moreover, a lower and an upper bound
	to the trace distance, can be given in terms of~fidelity:
	\begin{equation}
		1-\sqrt{F(\hat{\rho}_1,\hat{\rho}_2)}\leq \frac12 ||
		\hat{\rho}_1-\hat{\rho}_2||_1\leq
		\sqrt{1-F(\hat{\rho}_1,\hat{\rho}_2)}\,.
	\end{equation}
	By means of the above relationships, one can surmise that a high value of fidelity reflects
	having two neighbouring states (in the geometric sense) in the Hilbert space.
	
	\subsection{Bell-Nonclassicality vs. Fidelity }
	
	{
		Given the set of two-qubit density matrices $\Omega$ randomly generated according to the prescription in Section \ref{sec:chsh-random},
		we identify as a target state $\rho_T\in \Omega$ one having high purity and for which
		the value $\langle\mathcal{B}\rangle_{\rho_T}$ is close to the maximum quantum violation quantified by the Tsirelson's bound $\beta_Q= 2 \sqrt{2}$ \cite{Cirelson1980}, namely we have $1 - \frac{\mathcal{B}}{\beta_Q}~\simeq~0.0336.$ }
	
	In fact, let us suppose that we are in a feasible scenario in which it is possible to
	certify that this state gives its maximal violation of the CHSH--Bell inequality for a precise given set of measurements.
	
	At this stage, we assume that the alignment of the measurements settings can be trusted with arbitrary confidence
	(viz., the angles of the polarisers if dealing with experiments involving polarised photons,
	or the axis of the Stern--Gerlach experiments if dealing with spin$-1/2$ particles)
	while a lower level of trust can be assumed for the preparation of the state.
	For this reason, we can think of the real state as described by a randomly sampled density matrix in the neighbourhood of the target state.
	{Therefore, we study this situation assuming that a random bipartite state $\hat \rho\in F_\alpha\subset\Omega$ such that the event $F_\alpha$ is the set
		\begin{equation}
			F_\alpha=\{\rho\in \Omega_\alpha:\, \mathcal{F}(\rho,\rho_T)\ge\alpha\}.
	\end{equation}}
	\vspace{-12pt}
	
	{To tackle such a scenario, we generated three ensembles of random density matrices containing, respectively,
		$|\Omega_{0.75}|=2 \times 10^6,$ $|\Omega_{0.85}|= 1 \times 10^7,$ and $|\Omega_{0.95}|= 2 \times 10^8,$
		elements with the same procedure as in Section \ref{sec:appendix_random}.}
	{Later, we selected from each of the three ensembles
		a subset $F_\alpha$ containing $|F_\alpha|=\{4874, 1609, 110\}$
		states having a fidelity with the one we report in Appendix \ref{sec:appendix_target} greater than
		three different thresholds, i.e, $F_\alpha = \{0.75, 0.85, 0.95 \}.$
		It is worth noting that the cardinality of the three generated ensemble scales with the higher level
		of fidelity required. The stricter the requirement on fidelity is, the more difficult it is to randomly generate a state satisfying the closeness requirement.
		We refer to Appendix \ref{sec:appendix_tip_fid} for the typical distribution of the fidelity between two random states of dimension $d=4$.}
	
	Notice that the states generated to satisfy the properties that the reduce states are $\mathrm{Tr}_A \rho_T=\mathrm{Tr}_B \rho_T=\bm{1}/2$ and such that the real part of the coefficient is reflected with respect to the antidiagonal.
	In this sense, we can say that the state chosen are filtered, namely by applying local unitary $U_A, U_B$ the amount of entanglement does not vary in a filtered state
	$\rho_f=U_A \otimes U_B \rho U_A^\dag \otimes U_B^\dag$ but the reduced states are maximally mixed
	\cite{Guehne2009,Verstraete2003,Gisin1996,Leinaas2006}.

	For all the states, we computed the value of the operator defined in Equation~\eqref{eq:Bval}
	imposing the directions maximising the violation of the CHSH--Bell inequality for the target state $\hat \rho_T$ in Equation \eqref{eq:target} that is given in Equation \eqref{eq:target_p}.
	The histograms in Figure~\ref{fig:2} report the {probability density function} of observing a violation of local realism with fixed measurement
	settings when an inefficient state preparation is taken into account.
	We observe that the probability to violate the local realism by maintaining the measurement settings fixed increases
	with the closeness of the prepared state with the target one.
	Moreover, it seems that Bell nonclassicality is a more robust resource compared to entanglement or other indicators of nonclassicality, signalling
	that states close in the Hilbert space share a similar violation of the CHSH--Bell inequality despite the fact that strong constraints are imposed on the measurement settings \cite{fid1,fid2,fid3}.

	\subsection{Realising the Fidelity Constraint}
	To complement our analysis, we consider now a scenario in which the closeness assumption is relaxed while maintaining
	fixed measurement settings that allow the maximal violation of the CHSH--Bell inequality for $\rho_T.$
	Furthermore, in this case, we rely on a numerical analysis based on an ensemble of $M=10^6$ filtered random states $\{\rho_i \}_i=1^M$.
	Specifically, we computed the fidelity between any of those states with the target one $\rho_T$ and the value of the expression in
	\eqref{eq:Bval} with the measurement directions given in \eqref{eq:target_p}, namely $B_T=\langle\mathcal{B}\rangle_{\rho_i}.$
	
\begin{figure}[h]
		\includegraphics[width= 0.5 \textwidth]{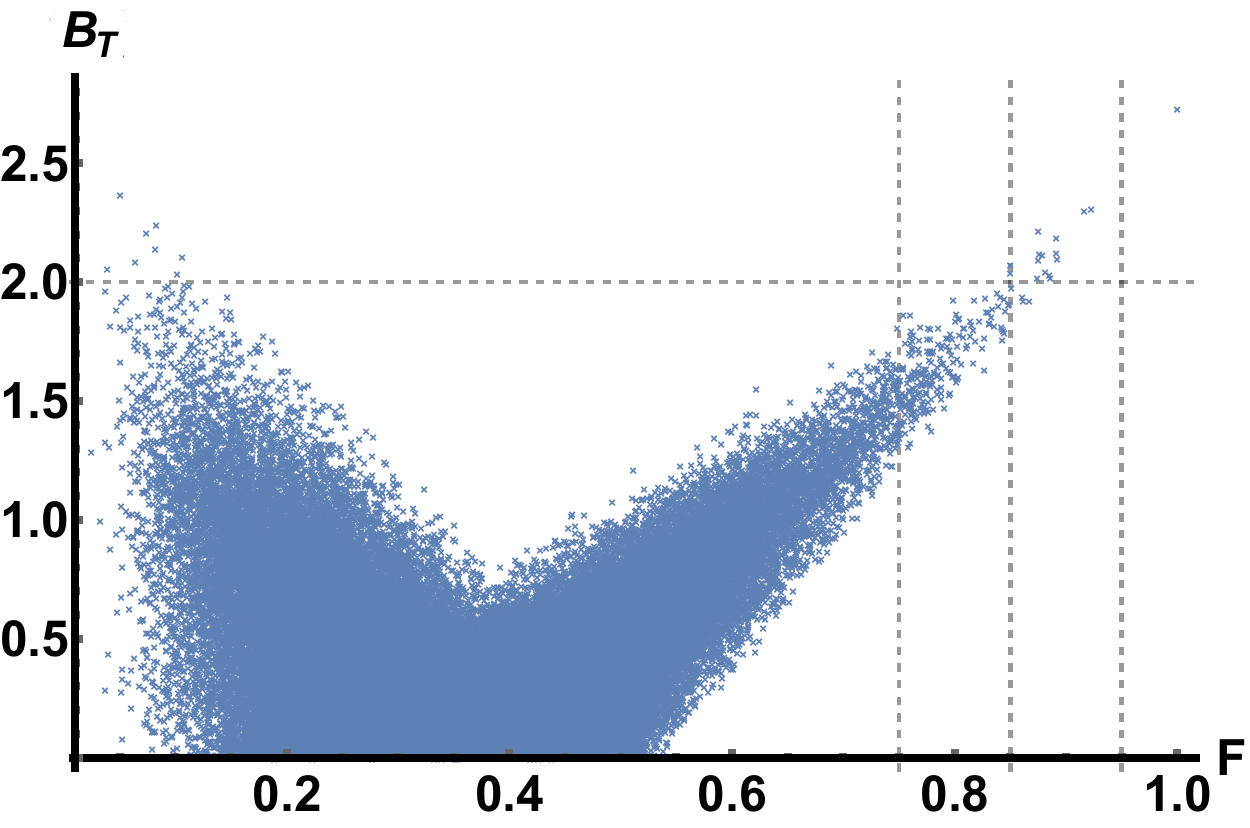}
		\caption{\label{fig:3}
			The distribution of $M=10^6$ random two-qubit filtered state in the plane given by the fidelity
			with the target state $\rho_T$ and the value of the CHSH--Bell function in the ``fixed settings'' scenario.
			Each cross ($\times$) in the plot corresponds to a random state.
			The dashed gray vertical line corresponds to threshold value for observing a nonclassical behaviour $\beta_C=2$,
			while the three horizontal dashed lines represent the three values of fidelity previously reported $F=\{0.75, 0.85, 0.95 \}$. }
	\end{figure}

	We report in Figure \ref{fig:3} how the value of the CHSH--Bell function distributes as a function of fidelity.
	It is evident that only few states having a fidelity higher than $0.85$
	show a violation of the classical bound $\beta_C=2,$ and it is expected if we fix
	the measurement apparatus on a single setting for any state.
	However, we observe a phenomenon not expected and worth investigating.
	The random states distribute in the plane along a \emph{J-type} curve, and
	few of them that should be really far in the Hilbert space from the target one
	(their fidelity is $(F<0.2)$) are violating the CHSH--Bell inequality.
	
	We surmise that there could be a certain symmetry in the ensemble of states generated with the procedure in
	Section \ref{sec:appendix_random}, probably related to the fact that the reshuffling operation
	on the unitary matrix gives a proper Wishart matrix only in the high dimension limit.
	Regardless, whether such a symmetry exists and how it relates to the choice of optimal measurement
	settings are beyond the scope of this paper and worthy of future investigation. 

	\section{Discussion and Future Perspectives}
	We studied a definition of \emph{continuity} as a notion of the \emph{robustness} of the bipartite mixed entangled state that violates the simplest Bell scenario. We conclude that the continuity does not hold, indeed, even if $\rho_T$ gives $B_T\in[2.72, 2.82]$ the family of state in a neighbourhood of $\rho_T$ with fidelity higher than $0.85$ might not certify a violation of Bell inequality.
	{However, the problem of whether a constraint of the fidelity threshold is enough to guarantee the detection of Bell nonclassicality measured via a ``fixed'' scenario situation optimised for a target state is still an open problem.}
	
	It means that despite the fact that the outcomes of incompatible measurements made by Alice and Bob sharing an entangled mixed state can exhibit stronger correlations than any resulting from an LHV model, a slight variation in the state in terms of fidelity can easily lose this strength and consequently admit correlation provide by a mere LHV model.
	
	In terms of quantum technologies, the implementation of devices that certifies the \emph{quantumness}, hence outperforming classical computational task is far to be trivial and requires rigorous self-test protocols to verify the quantum behaviour of some untrusted devices \cite{McKague2017,Coladangelo2018,Gigena2022}.
	
	In addition to that, we also observe that the generation of the mixed entangled states admits as reduced local states the maximally mixed state, and in this framework, our method can be also useful to certify the generation of a particular family of entangled states by defining a nonlocal witness that is also an entanglement witness\cite{Sarbicki2020,Lewenstein2000,Sarbickijpa_2020,Sarbicki2021}.
	
	Moreover, knowing the maximal classical and quantum bound for a more general Bell scenario \cite{Bae2018, Karczewski2022, Bernards2020}, for future research, one can extend the method we presented, which is also useful to test the criteria of nonclassicality and determine whether the trend of \emph{robustness} is a general property that does not depend on the particular qubit--qubit Bell scenario, but it can also be applied for different notions of nonclassicality as well as generalised noncontextuality~\cite{Catani2022,Catani2022a}. 

	\vspace{6pt}
	
	
	
	\section*{Acknowledgments}
	A.M. acknowledges the University of Catania and the University of Palermo for the hospitality and
	all the participants of the 14th Italian Quantum Information Science Conference IQIS22 held in Palermo for the inspiring conversations.
	\appendix

	\section{Target Bipartite State Considered}
	\label{sec:appendix_target}
	In this appendix, we report the explicit form of the random state we used as reference state in Section \ref{sec:fidelity}.
	
	\begin{equation}
		\label{eq:target}
		\rho_T=\left(\begin{array}{cccc}
			0.275 & 0.187-0.150 i &0.134+0.197 i &-0.24-0.079 i\\
			0.187+0.150 i &0.225 & -0.023+0.22 i &-0.134-0.197 i\\
			0.134-0.197 i &-0.023-0.22 i &0.225 & -0.187+0.150 i\\
			-0.24+0.079 i &-0.134+0.197 i &-0.187-0.150 i &0.275 \\
		\end{array}
		\right).
	\end{equation}
	
	The maximal violation of the CHSH--Bell inequality in Formula \eqref{eq:Bval} for this state is $\mathcal{B}=2.733$ and it is obtained considering the following
	values of parameters defining the two pairs of local observables:
	
	\begin{equation}
		\begin{split}
			\label{eq:target_p}
			&\theta_{A_1}\simeq 0.470262, \quad\phi_{A_1}\simeq 4.4278, \quad\theta_{B_1}\simeq 1.7903, \quad\phi_{B_1}\simeq 6.03714, \\
			&\theta_{A_2}\simeq 1.67123, \quad\phi_{A_2}\simeq 2.99268, \quad\theta_{B_2}\simeq 0.759639, \quad\phi_{B_2}\simeq 1.0833.
		\end{split}
	\end{equation}

	\section[\appendixname~\thesection]{Tipicality of the Fidelity with the Bipartite State Considered}
	\label{sec:appendix_tip_fid}
	In this appendix, we report the probability density functions of the fidelity that the target state $\rho_T$ in Equation (\ref{eq:target})
	has with random states generated according to the procedure outlined in Section \ref{sec:appendix_random} and according the Hilbert--Schmidt measure and the Bures metric~\mbox{\cite{zyczkowski1994random, zyczkowski2011generating}}.

\begin{figure}[h]
		\includegraphics[width= 0.5 \textwidth]{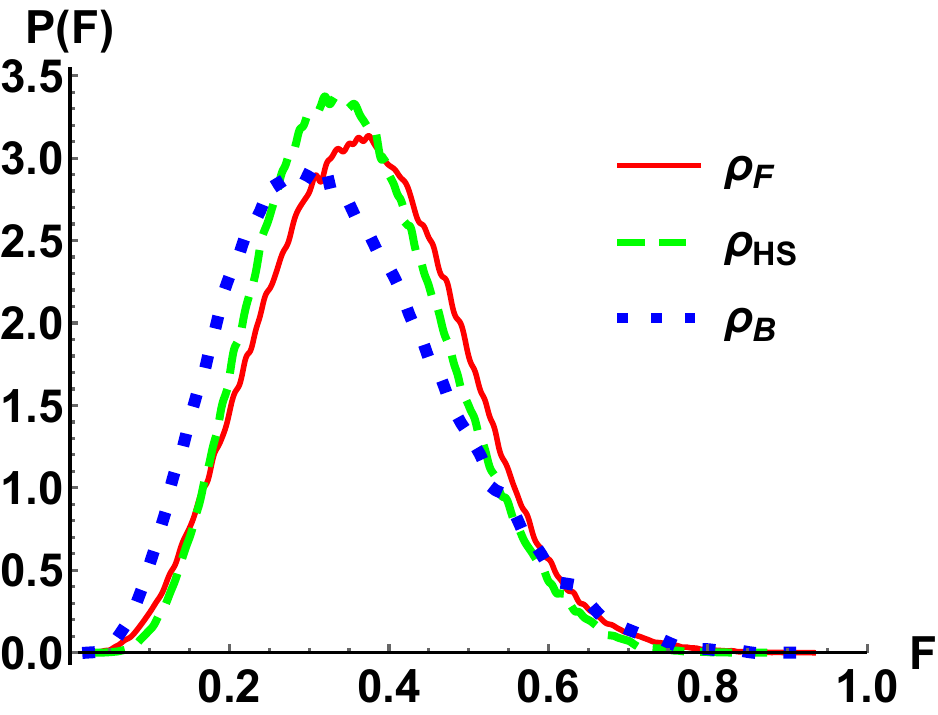}
		\caption{\label{fig:app} {The Probability Density Function $P(F)$ between the Randomly Chosen Target State $\rho_T$
				and: a random filtered mixed state $\rho_F$ (as the procedure in Section \ref{sec:appendix_random} (red solid line);
				a random mixed state $\rho_{HS}$ distributed according to the Hilbert--Schmidt measure (dashed green line);
				a random mixed state $\rho_{B}$ distributed according to the Bures measure (dotted blue line).
				The probability density functions are obtained numerically upon generation of three ensembles of $M=10^6$ random matrices~each.}}
	\end{figure}

	The results in Figure 
	\ref{fig:app} show that the fidelity distributions are typical for a
	quantum mixed state of dimension $d=4$ randomly generated and are in agreement with the results in \cite{Karol_fid}.
	We note that the distribution obtained considering filtered states does not match with any of the other two related to a proper metric.
	To the best of our knowledge, its behaviour is not known in the literature, and it opens the way for future investigations.
	
	\bibliography{CHSH_fid.bib}
\end{document}